# Definition of Reconnection Rate of Solar Flares Registered in 2011-2012 Years

A. T. Sarsembayeva

Physico-Technical Faculty, Theoretical and Nuclear Physics Department,
Al-Farabi Kazakh National University, Almaty 050040, Kazakhstan
aiganym@nucl.sci.hokudai.ac.jp

**Abstract**

Was defined reconnection rate of solar flares observed with the SOHO Michelson Doppler Imager (MDI). Measured physical parameters of 15 flares, such as the temporal scale, size and magnetic flux density. Estimated reconnection inflow velocity, coronal Alfven velocity, and reconnection rate using the observed values.

## I. Introduction

In the energy release process in solar flares, magnetic reconnection is generally considered to play a key role. The reconnection rate is an important quantity, because it puts critical restrictions on the reconnection model. It is defined as $M_A \equiv V_{in}/V_A$ in nondimensional form, where $V_{in}$ is the velocity of the reconnection inflow and $V_A$ is the Alfven velocity. It gives the normalized value of the reconnected flux per unit time. In spite of its importance, what determines the reconnection rate in flares is still a question [1].

In spite of its importance, what determines the reconnection rate in flares is still a question. In the steady reconnection model of Sweet [6] and Parker [3], the reconnection rate is $M_A = (\mathrm{Re}_m)^{-1/2}$, where $\mathrm{Re}_m = (V_A L/\eta)$ is the magnetic Reynolds number defined with the Alfven velocity, and $\eta$ is the magnetic diffusivity, $\eta \sim 10^4 (\frac{T}{10^6 K})^{-3/2} cm^2 s^{-1}$. In the solar corona, if the resistivity is attributed to Coulomb collisions [5], the typical value of $\text{Re}_m$ is $\text{Re}_m \sim 10^{14}$, which means that $M_A \sim 10^{-7}$ and the estimated timescale of the flare is about 1 yr. This is, of course, too slow to explain flares whose timescales are about $10^2 - 10^3 s$. On the other hand, Petschek [4] pointed out that the previous model lacks the effects of waves and suggested his model with $M_A \lesssim \pi / [8\ln(8\text{Re}_m)]$. The special feature of this model is that $M_A$ has a weak dependence on $\text{Re}_m$. In this model $M_A \lesssim 10^{-2}$ when $\text{Re}_m \sim 10^{14}$, and the estimated timescale is consistent with the observed value [2].

## II. Data analysis

The amount of energy released during a flare [1], $E_{flare}$, can be explained by the magnetic energy stored in the solar atmosphere,

$$E_{flare} \sim E_{mag} = \frac{B_{cor}^2}{8\pi} L^3 \quad (1)$$

where $L$ is the characteristic size of the flare and $B_{cor}$ is the characteristic magnetic flux density in the corona. Since the released magnetic energy balances the energy flowing into the reconnection region, we can describe the energy release rate as

$$\left|\frac{dE_{mag}}{dt}\right| \sim 2 \frac{B_{cor}^2}{4\pi} V_{in} L^2 \quad (2)$$

where $V_{in}$ is the inflow velocity of the plasma. Therefore, the time required for the energy inflow to supply the flare energy is estimated as

$$\tau_{flare} \sim E_{flare} \left(\left|\frac{dE_{mag}}{dt}\right|\right)^{-1} \sim \frac{L}{4V_{in}} \quad (3)$$

and this should be the timescale of the flare. Using this timescale, we can estimate the inflow velocity $V_{in}$ as

$$V_{in} \sim \frac{L}{4\tau_{flare}} \quad (4)$$

To evaluate the reconnection rate in nondimensional form, $M_A \equiv \dfrac{V_{in}}{V_A}$, we must estimate the Alfven velocity in the inflow region: $V_A = \dfrac{B_{cor}}{(4\pi\rho)^{1/2}}$. Hence, if we measure the coronal density $\rho$, the spatial scale of the flare $L$, the magnetic flux density in the corona $B_{cor}$, and the timescale of flares $\tau_{flare}$, we can calculate inflow velocity $V_{in}$, Alfven velocity $V_A$, and reconnection rate $M_A$.

## III. Results

Using the above described method, we analyzed 15 flares that have been registered in 2011-2012 years. Examined the dependence of the reconnection rate from GOES class of solar flares. Figure 1 shows the dependence of the reconnection rate from GOES class.

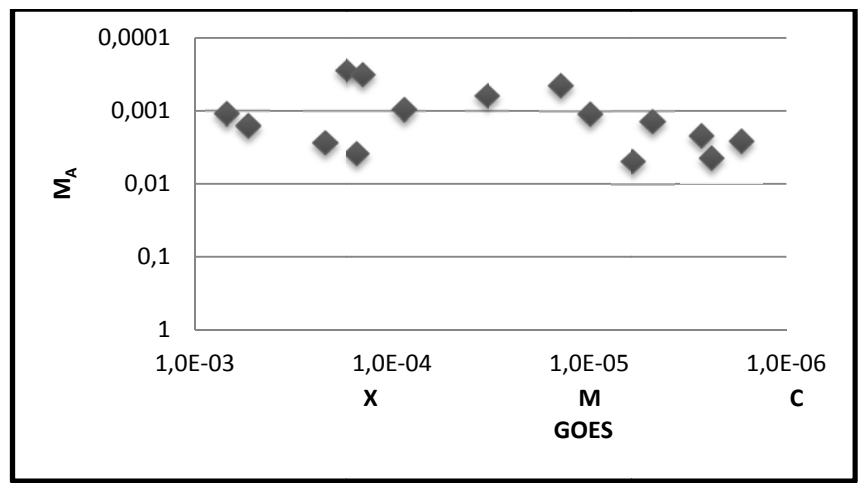

Fig. 1. Reconnection rate $M_A$ plotted against the GOES class of each flare

## IV. Conclusion

The values of reconnection rate are distributed in the range from $10^{-4} - 10^{-3}$. Here, the value of the reconnection rate decreases as the GOES class increases. The value of the reconnection rate obtained in this study is within 1 order of magnitude from the predicted maximum value of the Petschek model.

## References

[1]   K. Nagashima, Statistical study of the reconnection rate in solar flares observed with Yohkoh SXT, ApJ, 647, 654, 2006.
[2]   S. Nitta, ApJ, 610, 1117, 2004.
[3]   E. N. Parker, J. Geophys. Res., 62, 509, 1957.
[4]   H.E. Petschek, in The Physics of Solar Flares, ed. W. N. Hess (Washington: NASA), 425, 1964.
[5]   L. Spitzer, Physics of Fully Ionized Gases (New York: Interscience), 1956.
[6]   P.A. Sweet, in IAU Symp. 6, Electromagnetic Phenomena in Cosmical Physics, ed. B. Lehnert (Cambridge: Cambridge Univ. Press), 123, 1958.